\documentclass[prd,twocolumn,superscriptaddress,floatfix,nofootinbib]{revtex4-2}
\pdfoutput=1 
\usepackage[T1]{fontenc} 
\usepackage{amsmath,amssymb,amsfonts,amsthm}
\usepackage{graphicx}
\usepackage[usenames,dvipsnames]{color}
\usepackage{enumitem}
\usepackage{verbatim}
\usepackage[percent]{overpic}
\usepackage{rotating}
\usepackage{hyperref}
\usepackage{array}
\usepackage{mathtools}
\usepackage{lmodern}
\usepackage[normalem]{ulem}
\usepackage{braket}
\usepackage{tensor}
\usepackage{dsfont}
\usepackage{bm}
\usepackage{tikz}

\usepackage{mathrsfs}

\usetikzlibrary{decorations.pathmorphing}
\usepackage{CJKutf8}

\usetikzlibrary{positioning}
\usetikzlibrary{calc,through,backgrounds}

\theoremstyle{definition}
\newtheorem{definition}{Definition}[section]
\newtheorem{theorem}{Theorem}[section]

\newcommand{\kako}[1]{\left( #1 \right)}
\newcommand{\kagikako}[1]{\left[ #1 \right]}
\newcommand{\ts}[1]{ _{\text{#1}} }
\newcommand{\Bigkako}[1]{\Big( #1 \Big)}

\newcommand{\Bigkagikako}[1]{\Big[ #1 \Big]}

\newcommand{\mfd}{\mathcal{M}}

\newcommand{\rinda}{\hat a_{\omega,\bk_\perp}^{\text{R}}}
\newcommand{\rindap}{\hat a_{\omega',\bk'_\perp}^{\text{R}}}
\newcommand{\rindag}{\hat a_{\omega,\bk_\perp}^{\text{R}\dag}}
\newcommand{\rindagp}{\hat a_{\omega',\bk'_\perp}^{\text{R}\dag}}

\newcommand{\linda}{\hat a_{\omega,\bk_\perp}^{\text{L}}}
\newcommand{\lindap}{\hat a_{\omega',\bk'_\perp}^{\text{L}}}
\newcommand{\lindag}{\hat a_{\omega,\bk_\perp}^{\text{L}\dag}}
\newcommand{\lindagp}{\hat a_{\omega',\bk'_\perp}^{\text{L}\dag}}

\newcommand{\rindv}{v_{\omega,\bk_\perp}^{\text{R}}}

\newcommand{\lindv}{v_{\omega,\bk_\perp}^{\text{L}}}

\newcommand{\unruhw}[1]{w_{\omega,\bk_\perp}^{#1}}
\newcommand{\unruhws}[1]{w_{\omega,\bk_\perp}^{#1 *}}
\newcommand{\unrub}[1]{\hat b_{\omega,\bk_\perp}^{#1}}
\newcommand{\unrubd}[1]{\hat b_{\omega,\bk_\perp}^{#1 \dag}}
\newcommand{\unrubp}[1]{\hat b_{\omega',\bk_\perp'}^{#1}}
\newcommand{\unrubpd}[1]{\hat b_{\omega',\bk_\perp'}^{#1 \dag}}

\newcommand{\bsmr}[1]{\hat b({#1}_{\omega,\bk_\perp}^{\sigma *})}
\newcommand{\bsmrdag}[1]{\hat b^\dag({#1}_{\omega,\bk_\perp}^\sigma)}

\allowdisplaybreaks

\DeclareMathOperator{\Tr}{Tr}
\newcommand{\R}{\mathbb{R}}
\newcommand{\C}{\mathbb{C}}
\newcommand{\dd}{\text{d}}
\newcommand{\bk}{{\bm{k}}}
\newcommand{\bx}{{\bm{x}}}

\newcommand{\id}{\mathds{1}}
\newcommand{\sx}{\mathsf{x}}

\newcommand{\sz}{\mathsf{z}}

\newcommand{\ii}{\mathsf{i}}

\begin{document}

\title{Acceleration-induced radiation from a qudit particle detector model}


\author{Kensuke Gallock-Yoshimura}
\email{gallockyoshimura.kensuke@phys.kyushu-u.ac.jp} 

\affiliation{Department of Physics, Kyushu University, 744 Motooka, Nishi-Ku, Fukuoka 819-0395, Japan}

\author{Yuki Osawa}
 \email{osawa.yuki.e8@s.mail.nagoya-u.ac.jp}
\affiliation{Department of Physics, Nagoya University, Nagoya 464-8602, Japan}

\author{Yasusada Nambu}
\email[]{nambu.yasusada.e5@f.mail.nagoya-u.ac.jp}
\affiliation{Department of Physics, Nagoya University, Nagoya 464-8602, Japan}


\begin{abstract}
We nonperturbatively examine the emission rate of acceleration-induced radiation from a uniformly accelerated gapless qudit-type Unruh-DeWitt detector. 
We find that the emission rate can be written as Larmor’s formula multiplied by a factor that depends on the detector’s initial state. 
In particular, certain initial states of integer-spin detectors do not produce radiation. 
Although the appearance of Larmor's formula may suggest a classical phenomenon, we argue that the resulting radiation is fundamentally distinct from that of structureless classical sources, as it evolves into a multimode coherent state correlated with the detector's internal degree of freedom. 
Thus, gapless detectors cannot be treated as structureless sources, as previously proposed. 
\end{abstract}

\maketitle
\flushbottom

\section{Introduction}

Accelerated frames in quantum field theory offer invaluable insights into the fundamental mechanisms governing particle perception by observers in various states of motion. 
In particular, the Unruh effect states that an observer uniformly accelerated in the Minkowski vacuum perceives a thermal bath at a  temperature proportional to the acceleration \cite{Unruh1979evaporation}.

A closely related phenomenon is \textit{acceleration-induced radiation}, which is the primary focus of this paper. 
In classical electrodynamics, an accelerating charge is known to emit electromagnetic radiation, often referred to as Larmor radiation. 
Since the discovery of the Unruh effect, researchers have investigated the relationship between the Unruh thermal bath and such radiation. 
Key questions include: Is the Unruh thermal bath responsible for Larmor radiation? 
Do quantum systems with internal structure that experience the Unruh effect also emit radiation akin to electrically charged particles? 
What can be said about the classical/quantum nature of this radiation?

Although there was initial debate on whether uniformly accelerated particles radiate (see e.g., \cite{Lin.Hu.Accelerated.2006, Crispino2008review}), it is now widely accepted that quantum systems and classical sources radiate in $(3+1)$-dimensional Minkowski spacetime. 
For example, it has been shown that the total emission rate of Larmor radiation emitted by a charge follows from the exchange of zero-energy Rindler photons in the Unruh thermal bath \cite{Higuchi.Bremssstrahlung.1992, Higuchi.Bremsstrahlung.1992.part2, Crispino2008review, Paithankar.Brems.2020, Portales.Vector.Unruh.2022, Vacalis.Larmor.2024}. 
Here, the term ``zero-energy Rindler photons'' refers to the electromagnetic modes with vanishing Rindler frequency. 
The same argument also holds for a uniformly accelerated scalar source coupled to a scalar field \cite{Higuchi.classical.field.1993, Ren.Radiation.1994, Landulfo.Larmor-Unruh.2019}. 
We note that, in these studies, the source is a \textit{structureless classical source} under acceleration and thereby, the resulting emitted radiation is \textit{classical} Larmor radiation.

Meanwhile, uniformly accelerated \textit{quantum systems}, such as qubits and harmonic oscillators, have also been examined in this context. 
These quantum systems are commonly known as Unruh-DeWitt (UDW) particle detectors \cite{Unruh1979evaporation, DeWitt1979}, and their internal degrees of freedom also come into play. 
In \cite{Lin.Hu.Accelerated.2006}, a uniformly accelerated harmonic oscillator-type UDW detector is investigated by solving the equation of motion exactly. 
The paper shows that the radiation caused by the detector's internal dynamics is a transient effect, which vanishes as the detector-field system reaches equilibrium (see also \cite{Higuchi.entanglement.2017, Iso.radiation.entanglement.2017}). 
Nevertheless, the net radiation that escapes to null infinity remains nonzero even in the steady state, but it is fully attributed to the external force responsible for accelerating the detector.

More recently, Cozzella \textit{et al.} considered a \textit{gapless} two-level UDW detector and its relationship to zero-energy Rindler modes in \cite{Cozzella.UDW.2020}. 
Gapless detectors enable nonperturbative analyses, which are increasingly employed in the study of relativistic quantum information \cite{Landulfo2016magnus1, Simidzija2018no-go, Cozzella.UDW.2020, Landulfo.RQC2.2021, Tjoa.nonperturbative.gaussian, Perche.closedform.2024, Landulfo.RQC3.2024, Tjoa2024interacting}. 
Within both perturbative and nonperturbative regimes, they demonstrated that, in the long-interaction limit, a uniformly accelerated gapless two-level system (i) emits Minkowski particles at a rate agreeing with Larmor's formula, (ii) does so regardless of the detector's initial state, and (iii) that these radiated Minkowski particles originate from zero-energy Rindler modes in the Unruh thermal bath. 
From these observations, they claimed that the gapless two-level UDW detector can be thought of as a structureless classical source emitting Larmor radiation.

In this paper, we show that the conclusion reached by Cozzella \textit{et al.} in \cite{Cozzella.UDW.2020} for two-level systems does not hold in general for gapless \textit{multilevel} UDW detectors. 
In fact, we argue that the radiation emanating from a gapless detector (including the two-level model in \cite{Cozzella.UDW.2020}) is correlated with the detector's internal degree of freedom and must be distinguished from the classical Larmor radiation produced by structureless classical sources. 
In particular, we adopt the spin-$j$ representations of $SU(2)$ to model a $d(=2j+1)$-level gapless UDW detector, and derive nonperturbatively the total emission rate of Minkowski particles from a uniformly accelerated qudit. 
We find that, while the emission rate is proportional to Larmor's formula, it \textit{does} depend on the detector's initial state. 
Specifically, half-integer spins always radiate, but integer spins may not radiate for certain initial states. 
These initial states correspond to dark states \cite{lambropoulos2007fundamentals}, which do not respond to a given interaction Hamiltonian. 
Furthermore, the radiation is generally entangled with the detector, a feature that is absent in classical systems. 
From these observations, we argue that gapless multilevel UDW detectors, including the two-level ones, cannot be treated as structureless classical sources, in contrast to the claim in \cite{Cozzella.UDW.2020}.


The paper is organized as follows. 
After reviewing the quantum scalar field in Rindler frames in Sec.~\ref{subsec:Rindler modes}, we introduce the gapless spin-$j$ UDW detector in Sec.~\ref{subsec:UDW}. 
We then nonperturbatively derive the emission rate of Minkowski particles from a uniformly accelerated spin-$j$ gapless UDW detector in Sec.~\ref{sec:nonperturb results}. 
We also derive the emission rate for a spin-1 (i.e., qutrit-type) UDW detector using a perturbative method in Sec.~\ref{sec:perturbative analysis}, followed by our conclusion in Sec.~\ref{sec:conclusion}. 
Throughout this paper, we denote spacetime points by $\sx$, and use natural units $k\ts{B}=c=\hbar=1$ as well as the mostly-plus metric signature convention $(-,+,+,+)$.

\section{Setup}

\subsection{Quantum scalar field and Rindler mode}\label{subsec:Rindler modes}

In this section, we review the quantum scalar field in an accelerated frame based on \cite{Crispino2008review, Landulfo.Larmor-Unruh.2019, Higuchi.entanglement.2017}.

\begin{figure}[t]
\centering
\includegraphics[width=\linewidth]{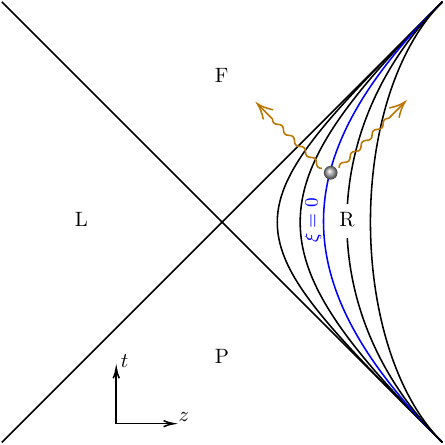}
\caption{Spacetime diagram depicting a uniformly accelerated quantum system in the right Rindler wedge (R). 
}
\label{fig:spacetime}
\end{figure}

\subsubsection{The right Rindler modes}

Consider a massless quantum scalar field in $(3+1)$-dimensional Minkowski spacetime. 
The field operator $\hat \phi(\sx)$ at a spacetime point $\sx$ satisfies the Klein-Gordon equation: 
\begin{align}
    (-\partial_t^2 + \nabla^2) \hat \phi(\sx)=0\,. \label{eq:KG eq}
\end{align}

Since we are interested in a uniformly accelerated detector, we adopt Rindler coordinates $(\tau, \xi, x,y)$ in the right Rindler wedge, $z>|t|$ (see Fig.~\ref{fig:spacetime}). 
The right Rindler coordinates are related to the Cartesian coordinates $(t,x,y,z)$ by 
\begin{align}
    t(\tau,\xi)
    &=
        \dfrac{e^{a\xi}}{a} \sinh(a\tau)\,,
    \quad
    z(\tau,\xi)
    =
        \dfrac{e^{a\xi}}{a} \cosh(a\tau)\,, \label{eq:trajectory}
\end{align}
where $\tau, \xi \in \R$, and $\tau$ and $a>0$ represent the proper time and proper acceleration of a uniformly accelerated observer along $\xi = 0$, respectively. 
From now on, the coordinates $(x,y)$ perpendicular to $(\tau, \xi)$ will be denoted by $\bx_\perp$. 
The metric covering the right Rindler wedge is then given by 
\begin{align}
    \dd s^2
    &=
        e^{2a\xi}(-\dd \tau^2 + \dd \xi^2)
        + \dd \bx_\perp^2\,. 
\end{align}
We note that $\partial_\tau$ is a Killing vector field in the right Rindler wedge.

The Klein-Gordon equation \eqref{eq:KG eq} reads 
\begin{align}
    (-\partial_\tau^2 + \partial_\xi^2 + e^{2a\xi} \partial_{\bx_\perp}^2) \hat \phi(\sx)=0\,,
\end{align}
and its solution can be written as 
\begin{align}
    \hat \phi(\sx\ts{R})
    &\equiv 
        \hat \phi(\tau, \xi, \bx_\perp) \notag \\
    &=
        \int_0^\infty \dd \omega
        \int_{\R^2} \dd^2 k_\perp
        \kagikako{
            \hat a_{\omega,\bk_\perp}^{\text{R}}
            v_{\omega,\bk_\perp}^{\text{R}}(\sx\ts{R})
            + \text{H.c.}
        }\,,
\end{align}
where 
\begin{align}
    v_{\omega,\bk_\perp}^{\text{R}}(\sx\ts{R})
    &=
        \sqrt{ \dfrac{ \sinh(\pi \omega/a) }{ 4\pi^4 a } }
        K_{\ii \omega/a}
        \kako{
            \dfrac{ |\bk_\perp| e^{a\xi} }{a}
        } \notag \\
        &\quad\times 
        e^{ -\ii \omega\tau + \ii \bk_\perp \cdot \bx_\perp }
\end{align}
is the right Rindler mode function. 
Here, $K_{\ii \nu} (x)$ is the modified Bessel function of the second kind, and if $\nu, x\in \R$ then $K_{\ii \nu}(x) \in \R$. 
Note that $v_{\omega,\bk_\perp}^{\text{R}}(\sx\ts{R})$ are positive frequency modes with respect to the timelike Killing vector field $\partial_\tau$. 
The creation ($\rindag$) and annihilation ($\rinda$) operators satisfy the commutation relations
\begin{subequations}
\begin{align}
    [\rinda, \rindagp]
    &=
        \delta(\omega - \omega') 
        \delta^{(2)}(\bk_\perp - \bk_\perp')\,, \\
    [\rinda, \rindap]
    &=
        [\rindag, \rindagp]
    =
        0\,,
\end{align}
\end{subequations}
and the right Rindler vacuum state $\ket{0\ts{R}}$ is defined as 
\begin{align}
    \rinda \ket{0\ts{R}}=0\,,
    \quad
    \forall \omega, \bk_\perp\,.
\end{align}

\subsubsection{The left Rindler modes}
The left Rindler wedge $-z > |t|$ can also be described by the coordinates $(\tilde \tau, \tilde \xi, \tilde \bx_\perp)$ with 
\begin{align}
    t(\tilde \tau, \tilde \xi)
    &=
        \dfrac{e^{a\tilde \xi}}{a} 
        \sinh(a\tilde \tau)\,,
    \quad
    z(\tilde \tau, \tilde \xi)
    =
        -\dfrac{e^{a\tilde \xi}}{a} 
        \cosh(a\tilde \tau)\,,
\end{align}
and the field operator can be expanded as 
\begin{align}
    \hat \phi(\sx\ts{L})
    &\equiv 
        \hat \phi(\tilde \tau, \tilde \xi, \tilde \bx_\perp) \notag \\
    &=
        \int_0^\infty \dd \omega
        \int_{\R^2} \dd^2 k_\perp
        \kagikako{
            \linda
            \lindv(\sx\ts{L})
            + \text{H.c.}
        }\,,
\end{align}
where 
\begin{align}
    \lindv(\sx\ts{L})
    &=
        \sqrt{ \dfrac{ \sinh(\pi \omega/a) }{ 4\pi^4 a } }
        K_{\ii \omega/a}
        \kako{
            \dfrac{ |\bk_\perp| e^{a \tilde \xi} }{a}
        } \notag \\
        &\quad\times 
        e^{ -\ii \omega \tilde \tau + \ii \bk_\perp \cdot \tilde \bx_\perp }
\end{align}
is the left Rindler mode function. 
The left Rindler mode functions $\lindv$ vanish in the right Rindler wedge. 
The creation and annihilation operators obey the commutation relations, 
\begin{subequations}
\begin{align}
    [\linda, \lindagp]
    &=
        \delta(\omega - \omega') 
        \delta^{(2)}(\bk_\perp - \bk_\perp')\,, \\
    [\linda, \lindap]
    &=
        [\lindag, \lindagp]
    =
        0\,,
\end{align}
\end{subequations}
and the left Rindler vacuum $\ket{0\ts{L}}$ is defined as 
\begin{align}
    \linda \ket{0\ts{L}}=0\,,
    \quad
    \forall \omega, \bk_\perp\,.
\end{align}

\subsubsection{The Unruh modes}

One can relate the positive frequency modes with respect to the Minkowski time $t$ to the left and right Rindler modes. 
Such modes, denoted by $\{ \unruhw{1}, \unruhw{2} \}$, are known as the Unruh modes, and they are given by 
\begin{subequations}
\begin{align}
    \unruhw{1}
    &\coloneqq
        \dfrac{ 
            \rindv + e^{-\pi \omega/a } v_{\omega, -\bk_\perp}^{\text{L}*}
        }
        {
            \sqrt{ 1-e^{-2\pi \omega/a} }
        }\,, \\
    \unruhw{2}
    &\coloneqq
        \dfrac{ 
            \lindv + e^{-\pi \omega/a } v_{\omega, -\bk_\perp}^{\text{R}*}
        }
        {
            \sqrt{ 1-e^{-2\pi \omega/a} }
        }\,.
\end{align}\label{eq:Unruh modes}
\end{subequations}
It is known \cite{Crispino2008review} that $\{ \unruhw{\sigma}, \unruhws{\sigma} \}$, $\sigma\in \{1,2\}$ forms a complete orthonormal set of solutions to the Klein-Gordon equation \eqref{eq:KG eq} with respect to the Klein-Gordon product. 
The field operator can be expressed as 
\begin{align}
    \hat \phi(\sx)
    &=
        \sum_{\sigma=1}^2 
        \int_0^\infty \dd \omega 
        \int_{\R^2} \dd^2 k_\perp
        \kagikako{
            \unruhw{\sigma} (\sx)
            \unrub{\sigma}
            +
            \text{H.c.}
        },
\end{align}
where $\unrub{\sigma}$ is the annihilation operator. 
The commutation relations are 
\begin{subequations}
\begin{align}
    [\unrub{\sigma}, \unrubpd{\sigma'}]
    &=
        \delta_{\sigma, \sigma'}
        \delta(\omega - \omega')
        \delta^{(2)} (\bk_\perp - \bk_\perp')\,, \\
    [\unrub{\sigma}, \unrubp{\sigma'}]
    &=
        [\unrubd{\sigma}, \unrubpd{\sigma'}]
    =0\,,
\end{align}
\end{subequations}
and $\unrub{\sigma}$ annihilates the Minkowski vacuum $\ket{0\ts{M}}$: 
\begin{align}
    \unrub{\sigma} \ket{0\ts{M}}=0\,,
    \quad
    \forall \sigma, \omega, \bk_\perp\,.
\end{align}

\subsection{The spin-$j$ UDW detector model}\label{subsec:UDW}

\subsubsection{Covariant formulation of the spin-$j$ UDW detector}

Let us introduce the spin-$j$ UDW particle detector model coupled to a quantum scalar field. 
Here, we consider a single detector moving along an arbitrary timelike trajectory in a general $(n+1)$-dimensional globally hyperbolic curved spacetime. 
In the following sections, we apply this framework to scenarios involving uniform acceleration.

Consider the $SU(2)$ spin-$j$ angular momentum operators $ \hat J_x, \hat J_y, \hat J_z$, which satisfy the commutation relation $[ \hat J_x , \hat J_y ]=\ii \hat J_z$ together with its cyclic permutations in $x,y,z$. 
The free Hamiltonian of the spin-$j$ UDW model is given by 
\begin{align}
    \hat H\ts{s,0}
    &=
        \Omega (\hat J_z + j \id)\,, \label{eq:free Hamiltonian}
\end{align}
where $\Omega \geq 0$ is the energy gap between adjacent energy levels, and $j$ can be an integer or a half-integer. 
For a fixed $j$, the energy eigenstates are given by the Dicke basis $\{ \ket{j,m}\,|\,m=-j, -j+1, \ldots, j \}$, and the corresponding energy eigenvalue of $\hat H\ts{s,0}$ is $(j+m)\Omega$. 
In particular, the energy eigenvalue for $\ket{j,-j}$ is $0$.

In the Schr\"odinger picture, we take the interaction Hamiltonian of the form $\sim \hat J_x \otimes \hat \phi(\bx)$. 
Let us consider the interaction Hamiltonian for the smeared detector in the interaction picture. 
Let $\sz(\tau)$ be the spacetime trajectory of the center of mass of the spin-$j$ UDW detector, whose proper time is $\tau$. 
In Secs.~\ref{sec:nonperturb results} and \ref{sec:perturbative analysis}, $\sz(\tau)$ corresponds to a uniformly accelerating trajectory \eqref{eq:trajectory}. 
The covariant prescription of the UDW detector model \cite{Tales2020GRQO, EMM2021brokencovariance} tells us that the interaction Hamiltonian density in the interaction picture is 
\begin{align}
    \hat h\ts{I}(\sx)
    &=
        \dfrac{1}{2} \Lambda(\sx) 
        (\hat J_+ e^{\ii \Omega \tau(\sx)} + \hat J_- e^{-\ii \Omega \tau(\sx)})
        \otimes \hat \phi(\sx)\,, \label{eq:covariant Hamiltonian density}
\end{align}
where $\Lambda(\sx)$ is the spacetime smearing function, and $\hat J_\pm \coloneqq \hat J_x \pm \ii \hat J_y$ are the raising and lowering operators.

The Hamiltonian density \eqref{eq:covariant Hamiltonian density} is coordinate independent when it is integrated over a given spacetime. 
The unitary time-evolution operator is thus given by 
\begin{align}
    \hat U\ts{I} 
    &=
        \mathcal T 
        \exp 
        \kagikako{
            -\ii \int_\mfd \dd V_{g'}\,
            \hat h\ts{I}(\sx')
        }\,,
\end{align}
where $\mathcal T$ is the time-ordering symbol, $\mfd$ represents the spacetime manifold, and $\dd V_{g'} \coloneqq \sqrt{-g'} \dd^n x'$ is the invariant volume element with the determinant of the metric $g$.

Let us now express $\hat U\ts{I}$ in the detector's reference frame. 
It is useful to employ the Fermi-normal coordinates $\bar \sx= (\tau, \bar \bx)$ associated to the detector's center of mass. 
In the detector's frame, we assume that the spacetime smearing can be separated into the temporal and spatial smearing functions as $\Lambda(\bar \sx)= \lambda \chi(\tau) f(\bar \bx)$, where $\lambda \geq 0$ is the coupling constant, $\chi(\tau)$ is the switching function, and $f(\bar \bx) $ is the spatial smearing function. 
Then, the time-evolution operator reads 
\begin{align}
    \hat U\ts{I}
    &=
        \mathcal{T}_\tau 
        \exp 
        \kagikako{
            -\ii \dfrac{\lambda}{2} 
            \int_\R \dd \tau\,\chi(\tau ) 
            (\hat J_+ e^{\ii \Omega \tau} + \hat J_- e^{-\ii \Omega \tau})
            \otimes 
            \hat \phi(f)
        }\,, \label{eq:time evolution with Omega}
\end{align}
where $\mathcal T_\tau$ is the time-ordering symbol with respect to the proper time $\tau$ and 
\begin{align}
    \hat \phi(f)
    &\coloneqq
        \int_{\Sigma_\tau} \dd^n \bar x\,
        \sqrt{-\bar g}\,
        f(\bar \bx) 
        \hat \phi(\bar \sx)
\end{align}
is the spatially smeared field operator with $\Sigma_\tau$ being a spatial hypersurface at constant $\tau$ and $\bar g$ is the metric determinant in the Fermi-normal coordinates. 
We can express Eq.~\eqref{eq:time evolution with Omega} with the interaction Hamiltonian $\hat H\ts{I}(\tau)$ as a generator of time translation with respect to $\tau$ as 
\begin{align}
    \hat U\ts{I}
    &=
        \mathcal{T}_\tau 
        \exp 
        \kagikako{
            -\ii 
            \int_\R \dd \tau\,
            \hat H\ts{I}(\tau)
        }\,, \label{eq:time evolution with Hamiltonian}
\end{align}
where 
\begin{align}
    \hat H\ts{I}(\tau)
    \coloneqq
        \dfrac{\lambda}{2} 
        \chi(\tau ) 
            (\hat J_+ e^{\ii \Omega \tau} + \hat J_- e^{-\ii \Omega \tau})
            \otimes 
            \hat \phi(f)\,. \label{eq:interaction Hamiltonian}
\end{align}

\subsubsection{Gapless spin-$j$ UDW model}

Typically, one employs perturbation theory by assuming the coupling constant $\lambda$ in \eqref{eq:time evolution with Omega} is very small. 
In contrast, a nonperturbative approach can be adopted by considering special scenarios, such as choosing the switching function $\chi(\tau)$ to be a Dirac delta (i.e., the instantaneous coupling) \cite{Simidzija.Nonperturbative, Simidzija2018no-go,Tjoa.RQC, Diana.tripartite.delta.2023, Tjoa.nonperturbative.gaussian, Jose.train.delta.2024}, or by setting the energy gap $\Omega $ to zero. 
In this paper, we focus on the latter---``gapless spin-$j$ model''---and nonperturbatively analyze the radiation emitted by such a detector (see e.g., \cite{Landulfo2016magnus1, Cozzella.UDW.2020, Tjoa.nonperturbative.gaussian, Perche.closedform.2024}).

Let us set $\Omega = 0$ in \eqref{eq:covariant Hamiltonian density}. 
Then, $\hat J_+ e^{\ii \Omega \tau} + \hat J_- e^{-\ii \Omega \tau}= 2\hat J_x$ and thus the commutator of two Hamiltonian densities at different spacetime points reads \cite{Perche.closedform.2024}
\begin{align}
    [ \hat h\ts{I}(\sx), \hat h\ts{I}(\sx') ]
    &=
        \ii \Lambda(\sx) \Lambda(\sx')
        E(\sx,\sx') \hat J_x^2 \otimes \id_\phi\,,
\end{align}
where $E(\sx,\sx')\coloneqq -\ii [ \hat \phi(\sx), \hat \phi(\sx') ] \in \R$ is the causal propagator. 
One can easily verify that the higher-order nested commutators, such as $[\hat h\ts{I}(\sx),[\hat h\ts{I}(\sx'), \hat h\ts{I}(\sx'')]]$, vanish. 
This result leads to a simplified expression for $\hat U\ts{I}$ using the Magnus expansion.

Consider the time-evolution operator in terms of the Magnus expansion \cite{Blanes.Magnus.2009}: 
\begin{align}
    \hat U\ts{I}
    &=
        \exp
        \kako{
            \sum_{j=1}^\infty \hat{\mathcal U}_j
        }\,,
\end{align}
where 
\begin{subequations}
\begin{align}
    \hat{\mathcal U}_1
    &\coloneqq
        -\ii \int_\mfd \dd V_g\,
        \hat h\ts{I}(\sx)\,, \\
    \hat{\mathcal U}_2
    &\coloneqq
        -\dfrac{1}{2}
        \int_\mfd \dd V_g 
        \int_\mfd \dd V_{g'}\, \Theta(t - t')
        [ \hat h\ts{I}(\sx), \hat h\ts{I}(\sx') ]\,, 
\end{align}
\end{subequations}
and $\hat{\mathcal U}_{j\geq 3}$ contains higher-order nested commutators. 
Here, $\Theta(t-t')$ denotes the Heaviside step function in terms of an arbitrary time parameter $t$. 
Since $[\hat h\ts{I}(\sx),\hat h\ts{I}(\sx') ]$ is the only nonvanishing commutator, we obtain 
\begin{align}
    \hat U\ts{I}
    &=
        e^{\hat{\mathcal U_1}}
        e^{\hat{\mathcal U_2}} \label{eq:gapless unitary}
\end{align}
with 
\begin{subequations}
\begin{align}
    \hat{\mathcal U}_1
    &=
        -\ii \hat J_x \otimes \hat \phi(\Lambda)\,, \\
    \hat{\mathcal U}_2
    &=
         \dfrac{-\ii}{2} G\ts{R}(\Lambda, \Lambda) \hat J_x^2 \otimes \id_\phi\,, \label{eq:U2}
\end{align}
\end{subequations}
where 
\begin{align}
    \hat \phi(\Lambda)
    &\coloneqq
        \int_\mfd \dd V_g\,\Lambda(\sx) \hat \phi(\sx) \label{eq:spacetime smeared field ope}
\end{align}
is the spacetime-smeared field operator, and $G\ts{R}(\Lambda, \Lambda)$ is the spacetime-smeared retarded Green function given by 
\begin{align}
    G\ts{R}(\Lambda, \Lambda)
    &\coloneqq
        \int_\mfd \dd V_g
        \int_\mfd \dd V_{g'}
        \Lambda(\sx) \Lambda(\sx')
        G\ts{R}(\sx, \sx')\,,
\end{align}
where $G\ts{R}(\sx, \sx')\coloneqq \Theta(t-t') E(\sx, \sx')$ is the retarded Green function.

We comment on a few points. 
First, this formulation for the gapless spin-$j$ UDW detector holds in any $(n+1)$-dimensional globally hyperbolic spacetime. 
Although we considered the single-detector scenario, it can be generalized to the multipartite case by following, e.g., Ref.~\cite{Perche.closedform.2024}.

Second, our spin-$j$ UDW model includes the qubit case studied in \cite{Landulfo2016magnus1, Cozzella.UDW.2020, Tjoa.nonperturbative.gaussian, Perche.closedform.2024} by choosing $j=\frac12$. 
In this case, $\hat J_x = \hat \sigma_x/2$ and the exponential of the operator $\hat{\mathcal U}_2$ in \eqref{eq:gapless unitary} becomes a \textit{global} phase. 
However, for $j\geq 1$, $\hat{\mathcal{U}}_2$ contributes to the \textit{relative} phase as we will see later. 
This difference alters the acceleration radiation results compared to the qubit model previously analyzed in \cite{Cozzella.UDW.2020}.

Finally, the same formulation can be applied to the pure dephasing model \cite{Tjoa.nonperturbative.gaussian}. 
Instead of setting $\Omega=0$, the dephasing model considers the interaction Hamiltonian (in the Schr\"odinger picture) of the form $\hat H\ts{int} \sim \hat J_z \otimes \hat \phi$. 
Since the free Hamiltonian \eqref{eq:free Hamiltonian} and the interaction Hamiltonian commute, the Hamiltonian density operator in the interaction picture becomes $\hat h\ts{I}(\sx)=  \Lambda(\sx) \hat J_z \otimes \hat \phi(\sx)$. 
Consequently, the same procedure can be employed to derive the time-evolution operator $\hat U\ts{I}$. 
In this case, the spin operator $\hat J_x$ in $\hat{\mathcal{U}}_j$ is replaced by $\hat J_z$.

\section{Nonperturbative analysis}\label{sec:nonperturb results}

\subsection{Emission rate}\label{subsec:emission rate}

Let us now focus on the scenario of a single spin-$j$ UDW detector uniformly accelerated in the right Rindler wedge of $(3+1)$-dimensional Minkowski spacetime. 
Before delving into our main calculations, let us introduce some new notations.

Consider the spacetime-smeared field operator $\hat \phi(\Lambda)$ given in \eqref{eq:spacetime smeared field ope}. 
Expressing it in terms of Unruh modes $\{ \unruhw{1}, \unruhw{2} \}$ gives 
\begin{align}
    \hat \phi(\Lambda)
    &=
        \sum_{\sigma=1}^2
        \int_0^\infty \dd \omega
        \int_{\R^2} \dd^2 k_\perp
        \int_\mfd \dd V_g\,
        \Lambda(\sx) \unruhw{\sigma} 
        \unrub{\sigma} \notag \\
        &\quad
        + \text{H.c.}\,,
\end{align}
where $(\mfd,g)$ is now understood to be Minkowski spacetime. 
Assuming that the spacetime smearing function is real, $\Lambda(\sx) \in \R$, we introduce 
\begin{subequations}
\begin{align}
    \ii \Lambda_{\omega, \bk_\perp}^{\sigma}
    &\coloneqq
    \int_\mfd \dd V_g\,
        \Lambda(\sx) \unruhws{\sigma} (\sx)\,, \label{eq:Fourier smearing}\\
    -\ii \Lambda_{\omega, \bk_\perp}^{\sigma *}
    &\coloneqq
    \int_\mfd \dd V_g\,
        \Lambda(\sx) \unruhw{\sigma} (\sx)\,, 
\end{align}
\end{subequations}
and 
\begin{subequations}
\begin{align}
    \bsmr{\Lambda}
    &\coloneqq
        \sum_{\sigma=1}^2
        \int_0^\infty \dd \omega
        \int_{\R^2} \dd^2 k_\perp\,
        \Lambda_{\omega, \bk_\perp}^{\sigma *} 
        \unrub{\sigma}\,, \\
    \bsmrdag{\Lambda}
    &\coloneqq
        \sum_{\sigma=1}^2
        \int_0^\infty \dd \omega
        \int_{\R^2} \dd^2 k_\perp\,
        \Lambda_{\omega, \bk_\perp}^{\sigma} 
        \unrubd{\sigma}\,,
\end{align}
\end{subequations}
so that the spacetime-smeared field operator can be compactly expressed as 
\begin{align}
    -\ii \hat \phi(\Lambda)
    &=
        \bsmrdag{\Lambda}
        -
        \bsmr{\Lambda}\,.
\end{align}
Using this notation, the time-evolution operator \eqref{eq:gapless unitary} becomes 
\begin{align}
    \hat U\ts{I}
    &=
        \exp
        \kako{
            \hat J_x \otimes 
            \kagikako{
                \bsmrdag{\Lambda} - \bsmr{\Lambda}
            }
        } \notag \\
    &\quad
    \times 
    \exp 
    \kagikako{
         \dfrac{-\ii}{2} G\ts{R}(\Lambda, \Lambda) \hat J_x^2 \otimes \id_\phi
    }\,.
\end{align}
We note that the first exponential resembles the displacement operator $\hat D(\alpha)=e^{ \alpha \hat a^\dag - \alpha^* \hat a }$ for $\alpha \in \C$. 
Indeed, as we show below, this operator serves as a multimode displacement operator that produces Larmor-like radiation.\footnote{In fact, the elements of the Weyl algebra $e^{\ii \hat \phi(\Lambda)}$ take the form of displacement operators.}

Let us now calculate the final quantum state. 
We prepare the initial state of the entire system as 
\begin{align}
    \rho\ts{s,0}\otimes \ket{0\ts{M}}\bra{0\ts{M}}\,,
\end{align}
where $\ket{0\ts{M}}$ is the Minkowski vacuum and $\rho\ts{s,0}$ is the initial state of the detector. 
In particular, we assume that $\rho\ts{s,0}$ is an arbitrary mixed state of the form 
\begin{align}
    \rho\ts{s,0}
    &=
        \sum_l p_l \ket{\psi_l} \bra{\psi_l}\,,
    \quad
    \ket{\psi_l}
    =
        \sum_{m_x=-j}^j 
        \alpha^{(l)}_{m_x} \ket{j,m_x}\,, 
        \label{eq:qudit initial state 1}
\end{align}
where $p_l \in [0,1]$ is the probability associated with the state $\ket{\psi_l}$, $\alpha^{(l)}_{m_x} \in \C$ are the normalized superposition coefficients, and $m_x\in \{-j, -j+1, \ldots, j\}$ is the magnetic quantum number along the $x$-axis, satisfying $\hat J_x \ket{j,m_x}=m_x \ket{j,m_x}$.

Our goal is to derive an expression for the total emission rate $\Gamma\ts{em}$ of Minkowski particles. 
To this end, we first compute $\hat U\ts{I} \ket{\psi_l}\ket{0\ts{M}}$, which can be done straightforwardly as 
\begin{align}
    \hat U\ts{I} \ket{\psi_l}\ket{0\ts{M}}
    &= 
        \sum_{m_x=-j}^j \alpha^{(l)}_{m_x}
        e^{ -\ii G\ts{R}(\Lambda, \Lambda) m_x^2/2 }
        \ket{j,m_x} \notag \\
        &\quad
        \otimes \hat D(m_x \Lambda_{\omega, \bk_\perp}^\sigma) \ket{0\ts{M}}\,, \label{eq:time evolved pure state}
\end{align}
where 
\begin{align}
    &\hat D(m_x \Lambda_{\omega, \bk_\perp}^\sigma)
    \coloneqq \notag \\
    &
        \exp
        \kagikako{
            \sum_{\sigma=1}^2 
            \int_0^\infty \dd \omega 
            \int_{\R^2} \dd^2 k_\perp\,
            m_x \Lambda_{\omega, \bk_\perp}^\sigma \unrubd{\sigma}
            - \text{H.c.}
        }.
\end{align}
From \eqref{eq:time evolved pure state}, we see that (i) the detector acquires a relative phase that depends on $m_x^2$, (ii) the Minkowski vacuum evolves to a coherent state, and (iii) the detector is entangled with the field. 
For $j=1/2$, this relative phase is purely global, consistent with the qubit case of \cite{Cozzella.UDW.2020}. 
However, for $j\geq 1$, different $m_x^2$ values introduce nontrivial phases that alter the detector's state (e.g., $m_x\in \{ \pm 3/2, \pm 1/2 \}$ for $j=3/2$). 
Moreover, integer-$j$ detectors allow for $m_x=0$. 
The detector initially in $\ket{\psi_l}=\ket{j,m_x=0}$ emits no radiation. 
Such a state corresponds to a dark state, which does not respond to the interaction Hamiltonian.

By tracing out the detector degree of freedom, we obtain the final state of the field: 
\begin{align}
    \rho_\phi
    &=
        \Tr\ts{s}[ \hat U\ts{I} (\rho\ts{s,0} \otimes \ket{0\ts{M}} \bra{0\ts{M}}) \hat U\ts{I}^\dag ] \notag \\
    &=
        \sum_l p_l
        \sum_{m_x=-j}^j |\alpha_{m_x}^{(l)}|^2 \notag \\
        &\quad
        \times 
        \hat D(m_x \Lambda_{\omega, \bk_\perp}^\sigma) \ket{0\ts{M}} \bra{0\ts{M}} \hat D^\dag(m_x \Lambda_{\omega, \bk_\perp}^\sigma)\,.
\end{align}
Introducing the Minkowski number operator
\begin{align}
    \hat N
    &\coloneqq
        \sum_{\sigma=1}^2
        \int_0^\infty \dd \omega 
        \int_{\R^2} \dd^2 k_\perp\,
        \unrubd{\sigma}
        \unrub{\sigma}\,, \label{eq:Unruh number ope}
\end{align}
we find the total emitted Minkowski particles $\braket{\hat N}\coloneqq \Tr[ \rho_\phi \hat N ]$ to be 
\begin{align}
    \braket{\hat N}
    &=
        \sum_l p_l
        \sum_{m_x=-j}^j |\alpha_{m_x}^{(l)}|^2 \notag \\
        &\quad \times 
        \braket{0\ts{M}| \hat D^\dag(m_x \Lambda_{\omega, \bk_\perp}^\sigma) \hat N \hat D(m_x \Lambda_{\omega, \bk_\perp}^\sigma) |0\ts{M}} \notag \\
    &=
        \mathcal N 
        \sum_l p_l
        \sum_{m_x=-j}^j |\alpha_{m_x}^{(l)}|^2 m_x^2 \,,
         \label{eq:general particle number}
\end{align}
with the Minkowski particle contribution
\begin{align}
    \mathcal{N}
    &\coloneqq
        \sum_{\sigma'=1}^2
        \int_0^\infty \dd \omega' 
        \int_{\R^2} \dd^2 k_\perp'\,
        |\Lambda_{\omega',\bk_\perp'}^{\sigma'}|^2\,. \label{eq:The N}
\end{align}
Here we have used 
\begin{align}
    \unrubp{\sigma'} \hat D(m_x \Lambda_{\omega, \bk_\perp}^\sigma) \ket{0\ts{M}}
    &=
        m_x \Lambda_{\omega', \bk_\perp'}^{\sigma'} 
        \hat D(m_x \Lambda_{\omega, \bk_\perp}^\sigma)
        \ket{0\ts{M}}\,,
\end{align}
and $\hat D^\dag(m_x \Lambda_{\omega, \bk_\perp}^\sigma) \hat D(m_x \Lambda_{\omega, \bk_\perp}^\sigma)=\id$.

So far, we have only assumed uniform acceleration in the right Rindler wedge, making Eq.~\eqref{eq:general particle number} valid for any such spin-$j$ detector. 
To proceed, we specify the spacetime spearing function $\Lambda(\sx)$ and compute the integral in \eqref{eq:The N} following \cite{Landulfo.Larmor-Unruh.2019, Cozzella.UDW.2020}. 
Consider a pointlike detector traveling along $(\tau, \xi, \bx_\perp)=(\tau, 0, \bm 0)$ in the right Rindler coordinates. 
The smearing function is then  
\begin{align}
    \Lambda(\sx)
    &=
        \lambda \chi(\tau) \delta(\xi) \delta^{(2)}(\bx_\perp) \,.
\end{align}
Note that we have not specified the switching function $\chi(\tau)$ yet. 
Since the left Rindler modes $\lindv$ in \eqref{eq:Unruh modes} vanish in the right Rindler wedge, $\Lambda_{\omega, \bk_\perp}^\sigma$ defined in \eqref{eq:Fourier smearing} become
\begin{subequations}
\begin{align}
    \Lambda_{\omega, \bk_\perp}^{1}
    &=
        \dfrac{ -\ii \lambda \tilde \chi(\omega) }{ \sqrt{ 1- e^{-2\pi \omega/a} }} 
        \sqrt{ \dfrac{ \sinh(\pi \omega/a) }{ 4\pi^4 a } } 
        K_{\ii \omega/a}
        \kako{
            |\bk_\perp|/a
        }\,, \\
    \Lambda_{\omega, \bk_\perp}^{2}
    &=
        \dfrac{ -\ii \lambda \tilde \chi^*(\omega) e^{-\pi \omega/a} }{ \sqrt{ 1- e^{-2\pi \omega/a} }} 
        \sqrt{ \dfrac{ \sinh(\pi \omega/a) }{ 4\pi^4 a } } 
        K_{\ii \omega/a}
        \kako{
            |\bk_\perp|/a
        }\,,
\end{align}
\end{subequations}
where $\tilde \chi(\omega)$ is the Fourier transform of the switching function: 
\begin{align}
    \tilde \chi(\omega)
    &\coloneqq
        \int_\R \dd \tau\,
        \chi(\tau) e^{\ii \omega \tau}\,.
\end{align}
Therefore, $\mathcal{N}$ in \eqref{eq:The N} is 
\begin{align}
    \mathcal N
    &=
        \dfrac{\lambda^2}{4\pi^2}
        \int_0^\infty \dd \omega'\,
        \dfrac{ \omega' ( 1+ e^{-2\pi \omega'/a} ) }{1- e^{-2\pi \omega'/a}}
        |\tilde \chi(\omega')|^2\,,
\end{align}
where we have used the identity \cite{Cozzella.UDW.2020}
\begin{align}
    \int_{\R^2} \dd^2 k_\perp\,
    K_{\ii \omega/a}^2(|\bk_\perp|/a)
    =
        \dfrac{ \pi^2 a \omega }{ \sinh (\pi \omega/a) }\,.
\end{align}

We now turn to the production \textit{rate} of Minkowski particles, i.e., $\braket{\hat N}$ per unit time. 
To this end, let us focus on $|\tilde \chi(\omega')|^2$ in $\mathcal N$, 
\begin{align}
    |\tilde \chi(\omega')|^2
    &=
        \int_\R \dd \tau 
        \int_\R \dd \tau'\,
        \chi(\tau) \chi(\tau') e^{ -\ii \omega' (\tau - \tau') }\,, \notag 
\end{align}
and choose a rectangular switching function supported in $\tau \in [ \tau_0, \tau\ts{f} ]$, where `f' stands for the final time. 
We then change integration variables $\tau, \tau'$ in such a way that, for $\tau'<\tau$, set $u\equiv \tau$, $s\equiv \tau - \tau'$, and for $\tau< \tau'$, set $u\equiv \tau'$, $s\equiv \tau' -\tau$ \cite{Schlicht.Fermi.Walker}. 
Rewriting the integral in these new variables yields 
\begin{align}
    |\tilde \chi(\omega')|^2
    &=
        \int_{\tau_0}^{\tau\ts{f}} \dd u\,
        2 \text{Re}\int_0^{u-\tau_0} \dd s\,e^{\ii \omega' s}\,. \label{eq:switching change var}
\end{align}
The integral with respect to $s$ can be interpreted as the ``rate'' at time $u$, and integrating this rate from $u=\tau_0$ to $u=\tau\ts{f}$ gives the total. 
Moreover, if we are interested in the long-interaction limit, we simply set $\tau_0 \to -\infty$, in which case the ``rate part'' becomes 
\begin{align}
    2 \text{Re}\int_0^{\infty} \dd s\,e^{\ii \omega' s}
    &=
        2\pi \delta(\omega')\,. \label{eq:long interaction}
\end{align}
The quantity $\mathcal N$ is then 
\begin{align}
    \mathcal{N}
    &=
        \int_{-\infty}^{\tau\ts{f}} \dd u\,
        \dfrac{\lambda^2}{4\pi^2}
        \int_0^\infty \dd \omega'\,
        \dfrac{ \omega' ( 1+ e^{-2\pi \omega'/a} ) }{1- e^{-2\pi \omega'/a}}
        2\pi \delta(\omega') \notag \\
    &=
        \int_{-\infty}^{\tau\ts{f}} \dd u\,
        \dfrac{\lambda^2 a}{4\pi^2}\,, \label{eq:mathcal N}
\end{align}
where we used the convention that $\int_0^\infty \dd \omega'\,f(\omega')\delta(\omega')=f(0)/2$.

Therefore, we have reached our conclusion that the emission rate $\Gamma\ts{em}$ of Minkowski particles from a uniformly accelerated spin-$j$ gapless UDW detector in the long-interaction limit is 
\begin{align}
    &\braket{ \hat N}
    =
        \int_{-\infty}^{\tau\ts{f}} \dd u\,
        \Gamma\ts{em}\,, \notag \\
    &\Gamma\ts{em}
    =
        \dfrac{\lambda^2 a}{4\pi^2}
        \sum_l p_l
        \sum_{m_x=-j}^j |\alpha_{m_x}^{(l)}|^2 m_x^2\,. \label{eq:rate result}
\end{align}

\subsection{Remarks}

\textit{State dependence of the emission rate:}
Our final result \eqref{eq:rate result} tells us that the emission rate generally depends on the initial quantum state of the detector, even for the gapless model. 
As a special case, if we choose the qubit model by setting $j=1/2$ (or simply replace $\hat J_x$ in the Hamiltonian with $\hat \sigma_x$) then $\Gamma\ts{em}$ is independent of the initial state as claimed in \cite{Cozzella.UDW.2020}: 
\begin{align}
    \Gamma\ts{em}
    &=
        \dfrac{\lambda^2 a}{16 \pi^2}\,, \label{eq:qubit emission rate}
\end{align}
where we used the fact that $\sum_{m_x} |\alpha_{m_x}^{(l)}|^2=1$ and $\sum_l p_l=1$. 
However, this is not the case for a general qudit detector, as the $m_x^2$ takes various values. 
For example, consider a qutrit detector $j=1$, in which case we have $m_x \in \{ 0, \pm 1 \}$. 
The emission rate in this case reads 
\begin{align}
    \Gamma\ts{em}
    &=
        \dfrac{\lambda^2 a}{4 \pi^2}
        \kako{
            1
            - \sum_l p_l |\alpha_{0}^{(l)}|^2
        }
        \leq \dfrac{\lambda^2 a}{4 \pi^2}\,,
\end{align}
with the equality if and only if the initial state does not contain $\ket{j,m_x=0}$. 
On the other hand, if the initial state is the pure state $\ket{j,m_x=0}$ then there will be no radiation emitted from the accelerated qutrit.

\textit{Correlation between detector and radiation:} 
Let us examine the correlation between the qudit and the emitted radiation. 
For simplicity, we rewrite the qudit's initial state \eqref{eq:qudit initial state 1} as 
\begin{align}
    \rho\ts{s,0}
    &=
        \sum_{m_x, m_x'=-j}^j \rho_{m_x, m_x'} 
        \ket{j,m_x} \bra{j,m_x'}\,,
\end{align}
where $\rho_{m_x, m_x'} \in \C$. 
Then, the emission rate $\Gamma\ts{em}$ in \eqref{eq:rate result} and the final states of the qudit, $\rho\ts{s}$, and the field, $\rho_\phi$, can be rewritten as  
\begin{subequations}
\begin{align}
    &\Gamma\ts{em}
    =
        \dfrac{\lambda^2 a}{4\pi^2}
        \sum_{m_x=-j}^j m_x^2 \rho_{m_x, m_x}\,, \label{eq:rate result another version} \\
    &\rho\ts{s}
    =
        \sum_{m_x, m_x'} \rho_{m_x, m_x'}
        e^{ -\ii G\ts{R}(\Lambda, \Lambda) (m_x^2 - m_x^{\prime 2})/2 }
        e^{ -(m_x - m_x')^2 \mathcal{N}/2 } \notag \\
        &\qquad \times 
        \ket{j,m_x}\bra{j,m_x'}\,, \\
    &\rho_\phi
    =
        \sum_{m_x} \rho_{m_x, m_x}
        \hat D(m_x \Lambda_{\omega, \bk_\perp}^\sigma) \ket{0}\bra{0}
        \hat D^\dag(m_x \Lambda_{\omega, \bk_\perp}^\sigma)\,. \label{eq:field final state 2}
\end{align}
\end{subequations}
As an example for the qubit case $j=1/2$, the final density matrix $\rho\ts{s}$ in the basis $\{ \ket{+}, \ket{-} \} \equiv \{ \ket{j=1/2, m_x=1/2}, \ket{j=1/2, m_x=-1/2} \}$ reads 
\begin{align}
    \rho\ts{s,0}
    &=
        \begin{bmatrix}
            \rho_{++} & \rho_{+-} \\
            \rho_{+-}^* & \rho_{--} 
        \end{bmatrix}
    \to 
    \rho\ts{s}
    =
        \begin{bmatrix}
            \rho_{++} & \rho_{+-} e^{ -\mathcal{N}/2 } \\
            \rho_{+-}^* e^{ -\mathcal{N}/2 } & \rho_{--} 
        \end{bmatrix}. \notag 
\end{align}

Consider the case where the qudit is initially prepared in a diagonal density matrix in the eigenbasis of $\hat J_x$. 
In this setup, $\rho\ts{s}$ remains constant throughout the interaction, i.e., $\rho\ts{s}=\rho\ts{s,0}$. 
No quantum correlations arise between the qudit and the radiation; instead, the field's state $\rho_\phi$ simply carries the information $m_x$ of the qudit through classical correlation.

On the other hand, if the qudit's density matrix initially has off-diagonal elements $\rho_{m_x, m_x'}$ with $m_x \neq m_x'$, the qudit becomes entangled with the radiation.\footnote{This is reminiscent of a controlled-NOT (CNOT) gate, which transfers the control qubit's information to the target via $\ket{a,b}\mapsto \ket{a,a\oplus b}$, and can entangle the inputs in cases like $\ket{+,0} \mapsto (\ket{0,0}+\ket{1,1}/\sqrt{2}$. } 
Moreover, all the off-diagonal entries of the final state $\rho\ts{s}$ acquire a factor of the form $e^{-\mathcal N}$, where $\mathcal N$ is given by \eqref{eq:mathcal N}. 
Since $\mathcal{N}$ increases with the interaction duration $\tau\ts{f}$, the final state $\rho\ts{s}$ decoheres significantly, and the qudit strongly entangles with the emitted radiation.

We finally comment on the qubit case. 
Both our results and those in \cite{Cozzella.UDW.2020} tell us that the emission rate for a two-level system is independent of its initial state. 
Thus, one might think that this scenario mimics a uniformly accelerated structureless classical source. 
However, this is not the case because the qubit's internal degree of freedom and the field are correlated. 
Specifically, the field's final state $\rho_\phi$ in \eqref{eq:field final state 2} still depends on the qubit's internal degree of freedom $m_x=\pm 1/2$. 
While the number of Minkowski particles produced and the emission rate for two-level detectors do not necessarily capture the internal structure of the qubit \cite{Cozzella.UDW.2020}, this information is imprinted in the quantum field's correlation functions.

\textit{Larmor-like radiation:} 
The formula for the emission rate \eqref{eq:rate result} [equivalently \eqref{eq:rate result another version}] is written as the Larmor-like radiation part, $\lambda^2 a/(4\pi^2)$, multiplied by the terms that only depend on the detector's state. 
The prefactor $\lambda^2 a/(4\pi^2)$ corresponds to the Larmor-like radiation because it originates from the expectation value of the Minkowski particle number in the coherent state, which is the most-classical state of light. 
However, the radiation possesses the information about the qudit's internal degree of freedom, which is absent in radiation from structureless sources. 

\textit{Zero-energy Rindler modes:} 
It is typically claimed that the Larmor radiation originates from the Unruh effect. 
In particular, Refs.~\cite{Higuchi.Bremssstrahlung.1992, Higuchi.Bremsstrahlung.1992.part2, Ren.Radiation.1994, Landulfo.Larmor-Unruh.2019} state that the zero-energy Rindler particles are responsible for the Larmor radiation emanating from a structureless classical source. 
Here, the zero-energy Rindler particles refer to the Rindler modes with $\omega=0$. 
This is also the case for gapless UDW detectors \cite{Cozzella.UDW.2020}. 
If the detector interacts with the field for a sufficiently long time, then the emission rate depends only on the Rindler modes with $\omega=0$ as one can see from $\delta(\omega)$ in \eqref{eq:mathcal N}. 

\section{Perturbative analysis: qutrit case}\label{sec:perturbative analysis}

We further show the nontriviality of acceleration radiation from a qudit detector by employing the perturbative method. 
For simplicity, let us focus on the $SU(2)$ qutrit model ($j =1$) \cite{Lima.qutrit.2023}. 

Consider a pointlike qutrit UDW detector with $\Omega \neq 0$ uniformly accelerating along $(\xi, \bx_\perp)=(0, \bm 0)$. 
Instead of using the Magnus expansion, we apply the Dyson series expansion to $\hat U\ts{I}$ in \eqref{eq:time evolution with Hamiltonian} as 
\begin{align}
    \hat U\ts{I}
    &=
        \id + \hat U\ts{I}^{(1)} + \hat U\ts{I}^{(2)} + \ldots\,,
\end{align}
where 
\begin{subequations}
\begin{align}
    \hat U\ts{I}^{(1)}
    &\coloneqq
        -\ii \int_{\R} \dd \tau\,
        \hat H\ts{I}(\tau)\,, \\
    \hat U\ts{I}^{(2)}
    &\coloneqq
        -\int_{\R} \dd \tau
        \int_{-\infty}^\tau \dd \tau'\,
        \hat H\ts{I}(\tau) \hat H\ts{I}(\tau')\,,
\end{align}
\end{subequations}
with 
\begin{align}
    \hat H\ts{I}(\tau)
    &=
        \dfrac{\lambda}{2} 
        \chi(\tau ) 
            (\hat J_+ e^{\ii \Omega \tau} + \hat J_- e^{-\ii \Omega \tau})
            \otimes 
            \hat \phi(\sx(\tau))\,.
\end{align}
Since we are using the qutrit model, the detector has three quantum states $\ket{j=1, m}, m\in \{ 0,\pm1 \}$, which are the eigenstates of $\hat J_z$. 
In what follows, we will simply write $\ket{j=1,m} \equiv \ket{m}$, and we remind ourselves that the energy eigenvalue of each state $\{ \ket{1}, \ket{0}, \ket{-1} \}$ is $\{ 2\Omega, \Omega, 0 \}$, respectively. 
Moreover, $\hat J_\pm$ are the raising and lowering operators satisfying 
\begin{align}
    \hat J_+ \ket{-1}
    &=
        \sqrt{2} \ket{0}\,,
    \quad
    \hat J_+ \ket{0}
    =
        \sqrt{2} \ket{1}\,, \notag \\
    \hat J_- \ket{1}
    &=
        \sqrt{2} \ket{0}\,,
    \quad
    \hat J_- \ket{0}
    =
        \sqrt{2} \ket{-1}\,,  \notag 
\end{align}
and otherwise $\hat J_\pm \ket{m}=0$.

The final density matrix of the quantum field $\rho_\phi$ can be obtained as 
\begin{align}
    \rho_\phi
    &=
        \Tr\ts{s}[ \hat U\ts{I} (\rho\ts{s,0} \otimes \rho_{\phi,0}) \hat U\ts{I}^\dag ]\,,
\end{align}
where the subscript `s' stands for the spin-$j$ UDW detector system. 
In the following, we assume that the quantum field is initially in the Minkowski vacuum $\rho_{\phi,0}=\ket{0\ts{M}}\bra{0\ts{M}}$, while the detector's initial state is arbitrary. 
In the qutrit case, $\rho\ts{s,0}$ in the basis $\{ \ket{1}, \ket{0}, \ket{-1} \}$ is expressed as 
\begin{align}
    \rho\ts{s,0}
    &=
        \begin{bmatrix}
            \rho_{11} & \rho_{12} & \rho_{13} \\
            \rho_{12}^* & \rho_{22} & \rho_{23} \\
            \rho_{13}^* & \rho_{23}^* & \rho_{33} 
        \end{bmatrix}\,,
    \quad
    \rho_{11} + \rho_{22} + \rho_{33}=1\,. \label{eq:initial density matrix}
\end{align}

To obtain the total emission rate $\Gamma\ts{em}$ of the Minkowski particles, we consider the expectation value of the (Unruh mode) number operator $\hat N$ given in \eqref{eq:Unruh number ope} in the state after the interaction. 
That is, 
\begin{widetext}
\begin{align}
    \braket{\hat N}
    &=
        \sum_{\sigma=1}^2 
        \int_0^\infty \dd \omega 
        \int_{\R^2} \dd^2 k_\perp 
        \Tr[ 
            \unrubd{\sigma} \unrub{\sigma} \hat U\ts{I}^{(1)} 
            (\rho\ts{s,0}\otimes \ket{0\ts{M}} \bra{0\ts{M}}) 
            \hat U\ts{I}^{(1)\dag}
        ] 
        + \mathcal{O}(\lambda^4)\,.
\end{align}
Note that this term is the sole contributor to $\braket{\hat N}$ at the leading order in $\lambda$.

Following a similar calculation in Sec.~\ref{sec:nonperturb results}, in particular using 
\begin{align}
    &\dfrac12 
    \sum_{\sigma=1}^2
    \int_0^\infty \dd \omega 
    \int_{\R^2} \dd^2 k_\perp
    \int_\mfd \dd V_g\,
    \Lambda(\sx) 
    \unruhws{\sigma}(\sx) e^{p \ii \Omega \tau} 
    \int_\mfd \dd V_{g'}\,
    \Lambda(\sx') 
    \unruhw{\sigma}(\sx') e^{q \ii \Omega \tau'} \notag \\
    &=
        \dfrac{\lambda^2}{8\pi^2}
        \int_0^\infty \dd \omega \,
        \dfrac{ \omega }{ 1-e^{ -2\pi \omega/a } }
        \Bigkagikako{
            \tilde \chi(\omega + p \Omega) \tilde \chi^*(\omega-q\Omega) 
            +
            e^{-2\pi \omega /a}
            \tilde \chi^*(\omega - p \Omega) \tilde \chi(\omega+q\Omega)
        }\,,
\end{align}
for $p,q \in \{ \pm 1 \}$, we obtain 
\begin{align}
    \braket{\hat N}
    &=
        \dfrac{\lambda^2}{8\pi^2}
        \int_0^\infty \dd \omega\,
        \dfrac{\omega}{ 1 - e^{ -2\pi \omega/a } } 
        \Bigkako{
            | \tilde \chi(\omega-\Omega) |^2
            [
                \rho_{11}
                + (1+ e^{-2\pi \omega/a}) \rho_{22}
                + e^{-2\pi \omega/a} \rho_{33}
            ] \notag \\
            &\quad
            +
            | \tilde \chi(\omega+\Omega) |^2
            [
                \rho_{33}
                + (1+ e^{-2\pi \omega/a}) \rho_{22}
                + e^{-2\pi \omega/a} \rho_{11}
            ] 
            + 2(1 + e^{-2\pi \omega/a}) 
            \text{Re}[ \rho_{13} \tilde \chi(\omega - \Omega) \tilde \chi^*(\omega + \Omega) ]
        }
        + \mathcal{O}(\lambda^4)\,. \label{eq:perturb particle number}
\end{align}
\end{widetext}

Observe that the off-diagonal term $\rho_{13}$ contributes to the emission when the qutrit is gapless $\Omega =0$ in \eqref{eq:perturb particle number}. 
By employing a rectangular switching function and applying the same argument as in Eqs.~\eqref{eq:switching change var} and \eqref{eq:long interaction}, the emission rate $\Gamma\ts{em}$ is related to $\braket{\hat N}$ as 
\begin{align}
    \braket{\hat N}
    &=
        \int_{-\infty}^{\tau\ts{f}} 
        \dd u\,\Gamma\ts{em}
        + \mathcal{O}(\lambda^4)\,;  \notag \\
    \Gamma\ts{em}
    &=
        \dfrac{\lambda^2 a}{4\pi^2}
        \kako{
            \dfrac{ 1+ \rho_{22} }{2} + \text{Re}[\rho_{13}]
        }\,, \label{eq:pert emission rate gapless}
\end{align}
which depends solely on the zero-energy Rindler modes. 
Recall that, for a gapless two-level detector, the emission rate of acceleration radiation is independent of the detector's initial state \cite{Cozzella.UDW.2020}. 
However, for gapless qutrit detectors (and qudit detectors in general) the emission rate depends on both the diagonal and off-diagonal elements of the initial density matrix $\rho\ts{s,0}$. 
This implies that a qudit with $\Omega=0$ cannot be considered a ``structureless scalar source'' in general. 
Moreover, it is possible for the number of radiated particles to vanish if a specific initial state $\rho\ts{s,0}$ is chosen. 
To illustrate this, consider the initial state $(\ket{1}-\ket{-1})/\sqrt{2}$, which corresponds to $\ket{0_x}$ in the $\hat J_x$ eigenbasis. 
For this state, we have $\rho_{22}=0$ and $\rho_{13}=-1/2$, leading to $\Gamma\ts{em}=0$. 
This is consistent with the nonperturbative analysis discussed in the previous section.

We also note that, for $\Omega \neq 0$, the Rindler modes with $\omega = \Omega$ contribute to $\braket{\hat N}$ in the long-interaction limit. 
By setting $\tau\ts{f}=\infty$ for simplicity,\footnote{If $\tau\ts{f} < \infty$, one must take into account the contribution from the last term in \eqref{eq:perturb particle number}, which depends on $\tau\ts{f}$ and $\Omega$. Taking the limit $\Omega \to 0$ should recover the gapless result in \eqref{eq:pert emission rate gapless}.} a similar calculation leads to 
\begin{align}
    \Gamma\ts{em}
    &=
        \dfrac{\lambda^2}{4\pi}
        \dfrac{ \Omega }{ 1 - e^{ -2\pi \Omega/a } } \notag \\
        &\quad\times 
        \kagikako{
            \rho_{11} 
            + (1 + e^{ -2\pi \Omega/a }) \rho_{22}
            + e^{ -2\pi \Omega/a } \rho_{33}
        }\,. \label{eq:number perturb gapped}
\end{align}
Observe that the emission rate is nonzero and depends only on the diagonal elements in the initial density matrix \eqref{eq:initial density matrix} of the detector. 
In addition, unlike the gapless detector model, the radiation consists of Rindler modes with energy $\omega= \Omega$.


\section{Conclusion}\label{sec:conclusion}
In this paper, we examined a gapless qudit UDW detector model undergoing uniform acceleration and derived, nonperturbatively, the emission rate of Minkowski particles emitted by the detector. 
The qudit is represented by an $SU(2)$ spin-$j$ detector model coupled to a massless quantum scalar field. 
Regardless of the interaction region (i.e., spacetime smearing), the quantum field evolves into a multimode coherent state correlated with the qudit. 
For a pointlike detector with a long-duration switching function, the emission rate is expressed as the product of the Larmor formula and terms dependent on the detector's initial state. 
Although the qubit model yields results independent of the detector's initial state \cite{Cozzella.UDW.2020}, this independence does not hold for $j\geq 1$. 
In fact, by choosing a specific initial state for the detector, one can entirely prevent radiation. 
Additionally, the results derived in this paper are valid for the pure dephasing model, where the energy gap is nonzero but no bit flips occur.

We also used a perturbative method to compute the emission rate for a qutrit with a nonzero energy gap. 
In this case, the emission rate remains nonvanishing for any initial state of the detector. 
However, for the zero energy gap, an off-diagonal coherence element in the initial density matrix emerges, allowing for the possibility of a zero emission rate for certain initial states.

Our findings tell us that uniformly accelerated gapless qudit-type UDW detectors (including the two-level ones) emit radiation that is distinct from those emanated from structureless sources, as qudits correlate (classically or quantumly) with the radiation. 
While it is true that the emission rate in the two-level UDW detector case is independent of the detector's initial state, the quantum field possesses the information of the detector's internal degree of freedom, which is absent in the case of structureless sources. 
The only scenario in which accelerating quantum systems can be treated as a classical source (without taking the classical limit) is when they have no internal structure that the radiation can correlate with.



\acknowledgments{
K.G.Y. and Y. O. thank Adam Wilkinson, Leo Parry, and Shih-Yuin Lin for helpful discussion at the early stage of this project. 
K.G.Y. is supported by Grant-in-Aid for Research Activity Start-up (Grant No. JP24K22862). Y. N. is supported by JSPS KAKENHI (Grant No. JP23K25871) and MEXT KAKENHI Grant-in-Aid for Transformative Research Areas A “Extreme Universe” (Grant No. 24H00956).
Y.O. would like to take this opportunity to thank the “Nagoya University Interdisciplinary Frontier Fellowship” supported by Nagoya University and JST, the establishment of university fellowships towards the creation of science technology innovation, Grant Number JPMJFS2120.
}

\section*{Data availability}
No data were created or analyzed in this study.

\bibliography{ref}

\end{document}